\documentclass[a4paper,superscriptaddress,nofootinbib,12pt]{article}
\usepackage{amssymb}
\usepackage{eurosym}
\usepackage{amsfonts}
\usepackage{geometry}
\usepackage{bbm}
\usepackage{authblk}
\usepackage[T1]{fontenc}
\def\bd{\begin{displaymath}}\def\ed{\end{displaymath}}
\def\be{\begin{equation}}\def\ee{\end{equation}}
\def\bea{\begin{eqnarray}}\def\eea{\end{eqnarray}}
\def\ba{\begin{array}}\def\ea{\end{array}}
\def\bs{\begin{split}}\def\es{\end{split}}
\def\lb{\label}

\def\a{\alpha}\def\b{\beta}\def\d{\delta}
\def\f{\phi}\def\g{\gamma}
\def\k{\kappa}\def\m{\mu}\def\n{\nu}\def\q{\psi}\def\r{\rho}\def\s{\sigma}\def\t{\tau}\def\u{\upsilon}
\def\y{\eta}
\def\D{\Delta}
\def\diag{{\rm diag}}\def\mo{{-1}}\def\ha{{1\over 2}}\def\lra{\leftrightarrow}
\def\mn{{\mu\nu}}\def\bdot{\!\cdot\!}\def\de{\partial}\def\id{\equiv}
\def\poi{Poincar\'e }\def\kp{$\k$-\poi}\def\cor{commutation relations }
\def\ie{i.e.\ }

\def\hx{\hat x}\def\hp{\hat p}

\def\rhd{\triangleright}\def\ta{\tilde\a}\def\tb{\tilde\b}
\def\tM{{\tilde M}}\def\tx{{\tilde X}}\def\tp{{\tilde P}}\def\tH{{\tilde H}}\def\tr{{\tilde\r}}
\def\f{\varphi}\def\hX{\hat x}\def\hP{\hat p}\def\hH{\hat h}\def\hh{\hat h}
\def\cD{{\cal D}}\def\cA{\tilde A}\def\cB{\tilde B}\def\cC{{\cal C}}\def\cF{{\cal F}}

\def\rhd{\triangleright}\def\ot{\otimes}
\def\act{\rhd1}\def\ab{{\a\b}}\def\ij{{ij}}\def\bt{{\bar t}}\def\bs{{\bar s}}\def\hM{{\hat M}}
\def\kd{$\k$-deformed }

\def\PL#1{Phys.\ Lett.\ {\bf#1}}\def\CMP#1{Commun.\ Math.\ Phys.\ {\bf#1}}
\def\PRL#1{Phys.\ Rev.\ Lett.\ {\bf#1}}
\def\PR#1{Phys.\ Rev.\ {\bf#1}}\def\CQG#1{Class.\ Quantum Grav.\ {\bf#1}}

\def\JMP#1{J.\ Math.\ Phys.\ {\bf#1}}
\def\PRS#1{Proc.\ R. Soc.\ Lond.\ {\bf#1}}
 \def\IJMP#1{Int.\ J. Mod.\ Phys.\ {\bf #1}}
\def\MPL#1{Mod.\ Phys.\ Lett.\ {\bf #1}}

\def\JHEP#1{JHEP\ {\bf#1}}\def\JCAP#1{JCAP\ {\bf#1}}
\def\AdP#1{Annalen Phys.\ {\bf#1}}
\def\arx#1{{\tt arXiv:#1}}

\begin{document}
\begin{titlepage}
\title{Realizations and star-product of doubly $\kappa$-deformed Yang models}
\vskip80pt
\author[1]{T. Martini\'c-Bila\'c{\footnote{teamar@pmfst.hr}}}
\author[2]{S. Meljanac{\footnote{meljanac@irb.hr}}}
\author[3,4]{S. Mignemi{\footnote{smignemi@unica.it}}}
\affil[1]{Faculty of Science, University of Split, Ru\dj era Bo\v skovi\'ca 33, 21000 Split, Croatia}
\affil[2]{Division of Theoretical Physics, Ru\dj er Bo\v skovi\'c Institute, \newline Bijeni\v cka cesta 54, 10002 Zagreb, Croatia}
\affil[3]{Dipartimento di Matematica e Informatica, Universit\`{a} di Cagliari, \newline via Ospedale 72, 09124 Cagliari, Italy}
\affil[4]{INFN, Sezione di Cagliari, Cittadella Universitaria, 09042 Monserrato, Italy}
\renewcommand\Affilfont{\itshape\small}
\maketitle
\thispagestyle{empty}

\begin{abstract}
The Yang algebra was proposed a long time ago as a generalization of the Snyder algebra to the case of curved background spacetime.
It includes as subalgebras both the Snyder and the de Sitter algebras and can therefore be viewed as a model of noncommutative curved
spacetime. A peculiarity with respect to standard models of noncommutative geometry is that it includes translation
and Lorentz generators, so that the definition of a Hopf algebra and the physical interpretation of the variables conjugated to the
primary ones is not trivial.
In this paper we consider the realizations of the Yang algebra and its $\k$-deformed generalization on an extended phase space and in
this way we are able to define a Hopf structure  and a twist.

\end{abstract}

\end{titlepage}

\section{Introduction}
The Yang model \cite{Yang} is an extension of the Snyder model \cite{Snyder} to the full phase space, obtained assuming that both
position and momentum operators do not commute among themselves.
It can therefore be interpreted as a noncommutative geometry defined on a spacetime of constant curvature.
Algebraically, it is based on an $so(1,5)$ algebra which includes the Lorentz generators and the position and momentum operators, together
with a further generator, necessary to close the algebra. Contrary to most models of noncommutative geometry, the action of the Lorentz
algebra on phase space is not deformed. It also enjoys an invariance under a (generalized) Born duality \cite{Born}.

The relevance of noncommutative geometries for Planck-scale physics has been originally highlighted in \cite{DFR} and since then the subject
has become fashionable. In particular the formalism of Hopf algebras has shown to be useful for the description of the physical implications
of the theory \cite{Maj}. Among these, particular attention have received the deformation of the symmetries of spacetime \cite{kP,AC,dsr} and the
applications to phenomenology and to the search for observable effects, especially in the context of Doubly Special Relativity \cite{AC,DSR}.
In particular, the introduction of a curved background can be useful in cosmological contexts, especially when considering
phenomenological effects on the propagation of photons from distant sources \cite{cosm}.

Thus, after a long oblivion, the Yang model was resumed in recent years. Some generalizations were presented by Khruschev and Leznov
(KL) \cite {KL} and in ref.~\cite{MS}, where an extension of the model that includes also the related Triply Special Relativity (TSR) theory
\cite{TSR} was introduced.
Further investigations concerning in particular its realizations on a canonical phase space have been recently performed in \cite{MM1,LMMP},
using the methods introduced in \cite{BM,GL} for the study of Snyder space. Other contributions to the study of Yang model are given in
\cite{ya}, while different models of noncommutative geometry in curved spaces can be found in \cite{nccs}.

Finally, in \cite{LMMPW} the Yang model has been further generalized by deforming the  $so(1,5)$ algebra to an  $so(1,5;g)$ algebra,
so that it also includes \kp deformations of the kind introduced in \cite{kP} for the standard \poi algebra, both in position and momentum
spaces.
Although from a mathematical point of view the two algebras are isomorphic, their physical interpretation is of course different, since,
for example, in $so(1,5;g)$ the \poi algebra is deformed.

In this paper, we study the explicit transformations leading from the generators of $so(1,5)$  to those of $so(1,5;g)$. This will allow us
to obtain a Hopf algebra structure for the Yang model, by defining a coproduct and a twist and then the star product. This will be done by
using the methods that have been proved successful in the case of Snyder \cite{BM} and \kd Snyder \cite{kSny} spaces.
In particular, exploiting the isomorphism of the Yang model to an orthogonal algebra, one can use the results of \cite{MMK}, where the
Hopf structure associated to orthogonal groups has been discussed in detail.

As shown in \cite{LMMP} a realization of the Yang model can be obtained on an extended phase space, that includes momenta canonically
conjugated to both the original position and momentum variables. Hence, in contrast with standard models of noncommutative geometry, in our
case a coproduct must be introduced also for the momentum variables.
The physical interpretation of this fact is presently under investigation.


\section{Generalized Yang models}
The Yang model \cite{Yang} is based on the Yang algebra, which is a Lie algebra generated by $\hx_i$, $\hp_i$, $\hM_\ij$ and $\hh$
and is isomorphic to the orthogonal algebra $so(1,5)$, $so(2,4)$ or $so(3,3)$, depending on the signature chosen for the metric
tensor \cite{MMM}. It is defined by the commutation
relations\footnote{Latin indices run from 0 to 3, Greek indices from 0 to 5. For definiteness in the following we
consider the $so(1,5)$ case with metric $\y_\ij=\diag(-1,1,1,1,1,1)$, but our considerations trivially extend to
the other cases. We use natural units, in particular we set $\hbar=1$.}
\bd
[\hx_i,\hx_j]=i\b^2\hM_\ij,\qquad[\hp_i,\hp_j]=i\a^2\hM_\ij,\qquad[\hx_i,\hp_j]=i\y_\ij\hh,
\ed
\be\lb{Yan}
[\hh,\hx_i]=i\b^2\hp_i,\qquad[\hh,\hp_i]=-i\a^2\hx_i,
\ee
where the $\hM_\ij$ generate the Lorentz algebra $so(3,1)$ and $\hx_i$, $\hp_i$ transform as vectors under the Lorentz
algebra, while $\hh$ is a scalar. The parameters $\a$ and $\b$ are constant with dimensions $[L]^\mo$ and $[M]^\mo$ respectively
and are usually identified with the square root of the cosmological constant and with the inverse of the Planck mass. 
In spite of the notation, $\a^2$ and
$\b^2$ can be taken negative leading to the $so(2,4)$ and $so(3,3)$ cases, which we shall not consider in detail here.

The operators $\hx_i$ and $\hp_i$
can be interpreted as the position and the momentum operators in a quantum phase space. The operator $\hh$ is necessary
to close the algebra and generates rotations in the $x$-$p$ hyperplane.
The Yang algebra satisfies the Born duality \cite{Born}, being invariant for $\a\lra\b$, $\hx_i\to-\hp_i$,  $\hp_i\to\hx_i$,
$\hM_\ij\lra\hM_\ij$, $\hh\lra\hh$.

Defining \be\lb{drY}
\hM_{i4}={\hx_i\over\b},\qquad\hM_{i5}={\hp_i\over\a},\qquad\hM_{45}={\hh\over\a\b}.
\ee
then the algebra (\ref{Yan}) can be put in the explicit $so(1,5)$ form
\be\lb{coY}
[\hM_\mn,\hM_{\r\s}]=i\big(\y_{\m\r}\hM_{\n\s}-\y_{\m\s}\hM_{\n\r}-\y_{\n\r}\hM_{\m\s}+\y_{\n\s}\hM_{\m\r}\big).
\ee
One can find alternative realizations of this algebra by defining linear combinations of the generators.
The most general new generators linear in $\hx_i$, $\hp_i$, $\hM_\ij$ are
\vfil\eject
\bea\lb{tran}
&&\tx_i=A(\cos\f\,\hx_i+{\b\over\a}\sin\f\,\hp_i)+\b a_k\hM_{ik},\cr
&&\cr
&&\tp_i=B(\cos\q\,\hp_i+{\a\over\b}\sin\q\,\hx_i)+\a b_k\hM_{ik},
\eea
with $\tM_\ij=\hM_\ij$. The parameters $A$, $B$, $\f$, $\q$, $a_i$, $b_i$ are dimensionless  with $AB\ne0$.
The transformations inverse to (\ref{tran}) are
\bea
\hx_i&=&{A^\mo\a\cos\q(\tx_i-\b a_j\tM_\ij)-B^\mo\b\sin\f(\tp_i-\a b_j\tM_\ij)\over\a\cos(\f+\q)}\cr
\hp_i&=&{B^\mo\b\cos\f(\tp_i-\a b_j\tM_\ij)-A^\mo\a\sin\q(\tx_i-\b a_j\tM_\ij)\over\b\cos(\f+\q)}
\eea

The new generators $\tx_i$ and $\tp_i$ generate a new class of Lie algebras isomorphic to the initial Yang algebra.
The new algebra generated by $\tx_i$, $\tp_i$, $\tM_\ij$ and $\tH$ is given by the following \cor
\bea\lb{newal}
[\tx_i,\tx_j]&=&i\left(\b^2\cA\tM_\ij+\b(a_i\tx_j-a_j\tx_i)\right),\cr
&&\cr
[\tp_i,\tp_j]&=&\left(\a^2\cB\tM_\ij+\a(b_i\tp_j-b_j\tp_i)\right),\cr
&&\cr
[\tx_i,\tp_j]&=&i\left(\y_\ij\tH+\a b_i\tx_j-\b a_j\tp_i+\a\b\tr\tM_\ij\right),\cr
&&\cr
[\tM_\ij,\tx_k]&=&i\Big(\y_{ik}\tx_j-\y_{jk}\tx_i+\b(a_i\tM_{kj}-a_j\tM_{ki})\Big),\cr
&&\cr
[\tM_\ij,\tp_k]&=&i\Big(\y_{ik}\tp_j-\y_{jk}\tp_i+\a(b_i\tM_{kj}-b_j\tM_{ki})\Big),\cr
&&\cr
[\tM_\ij,\tH]&=&i\Big(\a(b_j\tx_i-b_i\tx_j)-\b(a_j\tp_i-a_i\tp_j)\Big),\cr
&&\cr
[\tH,\tx_i]&=&i\left(\b^2\cA\tp_i-\a\b\tr\,\tx_i-\b a_i\tH\right),\cr
&&\cr
[\tH,\tp_i]&=&i\left(-\a^2\cB\tx_i+\a\b\tr\,\tp_i+\a b_i\tH\right),
\eea
where we have defined
\be
\tH=AB\cos(\f+\q)\hh+\b a\bdot\tp-\a b\bdot\tx-\a\b a_h b_k\hM_{hk},
\ee
\be
\tr=AB\r+a\bdot b,\qquad\cA=A^2+a^2,\qquad\cB=B^2+b^2,
\ee
with $\r=\sin(\f+\q)$. A generalized Born duality still holds for $\a\lra\b$, $a_i\to-b_i$, $b_i\to a_i$, $\cA\lra\cB$, $\tr\lra-\tr$,
$\tx_i\to-\tp_i$,  $\tp_i\to\tx_i$, $\tM_\ij\lra\tM_\ij$, $\tH\lra\tH$.

These \cor are of the kind introduced in \cite{LMMP}. They can be put in the form of as an $so(1,5;g)$ algebra
if one defines the generators $\tM_\mn$ as
\be\lb{dr}
\tM_{i4}={\tx_i\over\b},\qquad\tM_{i5}={\tp_i\over\a},\qquad\tM_{45}={\tH\over\a\b}.
\ee
They satisfy the algebra
\be\lb{cor}
[\tM_\mn,\tM_{\r\s}]=i\big(g_{\m\r}\tM_{\n\s}-g_{\m\s}\tM_{\n\r}-g_{\n\r}\tM_{\m\s}+g_{\n\s}\tM_{\m\r}\big),
\ee
with $g_\mn$ a symmetric matrix of the form
\begin{equation}
g_\mn=\left(
\begin{array}{cccccc}
-1&0&0&0&a_0&b_0\\
0&1&0&0&a_1&b_1\\
0&0&1&0&a_2&b_2\\
0&0&0&1&a_3&b_3\\
a_0&a_1&a_2&a_3&\cA&\tr\\
b_0&b_1&b_2&b_3&\tr&\cB
\end{array}
\right). \end{equation}
Notice that $\det g=A^2B^2\cos^2(\f+\q)$, hence one must require that $\cos(\f+\q)\ne0$, otherwise the matrix $g$
becomes singular.

The matrix $g_\mn$ can be reduced to a diagonal form by a transformation with a matrix $S$
such that $g=S\,\y\,S^T$. This matrix is defined up to multiplication by an orthogonal matrix $O$ such that $O\,\y\,O^T=\y$.
One can choose $S$ in a lower triangular form, as
\begin{equation}
S=\left(
\begin{array}{cccccc}
1&0&0&0&0&0\\
0&1&0&0&0&0\\
0&0&1&0&0&0\\
0&0&0&1&0&0\\
-a_0&a_1&a_2&a_3&\s&0\\
-b_0&b_1&b_2&b_3&\u&\t
\end{array}
\right),
\end{equation}
with
\be
\s=A,\qquad \u=B\sin(\f+\q),\qquad\t=B\cos(\f+\q),
\ee
and $\det S=\s\t=AB\cos(\f+\q)$.

Clearly, the generators $\hM_\mn$ such that $\tM_\mn=(S\,\hM\,S^T)_\mn$, satisfy the \cor (\ref{coY}) and can then be identified
with those defined in  (\ref{drY}).
The explicit relation between their components are then $\tM_\ij=\hM_\ij$, and
\bea
&&\tx_i={1\over\s}(\hX_i-a^k\hM_{ik}),\cr&&\cr
&&\tp_i={1\over\t}(\hP_i-b^k\hM_{ik})-{\u\over\s\t}(\hX_i-a^k\hM_{ik}),\cr&&\cr
&&\tH={1\over\s\t}(\hH+b^k\hX_k-a^k\hP_k+a^hb^k\hM_{hk}),
\eea
with inverse transformations
\bea
&&\hX_i=\s\tx_i+a^k\tM_{ik},\cr&&\cr
&&\hP_i=\t\tp_i+\t\tx_i+b^k\tM_{ik},\cr&&\cr
&&\hH=\s\t\,\tH+(\s b^k+\u a^k)\tx_k+\t a^k\tp_k+a^hb^k\tM_{hk}.
\eea
Note that if $a_k = b_k = 0$, then the KL model with $\ta$, $\tb$, $\tr$, $\tH$
is isomorphic to the original Yang model with $A$, $B$, $\r$, $h$ if
$\a^2 = \ta^2/B^2$, $\b^2 = \tb^2/A^2$ and $h =\tH/(AB\cos(\f+\q))$
where $A$, $B$ are real numbers and $A^2 B^2 > \tr^2$.

The same construction as (\ref{tran})-(\ref{newal}) holds also for Yang models isomorphic to $so(3,3)$.
For models isomorphic to $so(2,4)$, in eq.~(\ref{tran}) the trigonometric functions should be replaced by
the corresponding hyperbolic functions and the parameter $\r$ can be arbitrary; in particular, when
$\f=\q$ it follows that $\r=0$.

\section{Weyl representation of $so(1,5;g)$}
We shall now obtain representations for the \kd Yang model by using the previous relations between $so(1,5)$
and $so(1,5;g)$. Notice that from a four-dimensional point of view our primary fields are the generators $\tM_\ij$,
$\tx_i$, $\tp_i$ and $\tH$, to each of which we shall assign a conjugate momentum. The formalism is therefore analog to that
introduced for the extended Snyder model \cite{kSny}.

Consider then the generalized Heisenberg algebra
\be\lb{heis}
[x_\mn,x_{\r\s}]=[k^\mn,k^{\r\s}]=0,\qquad[x_\mn,k^{\r\s}]=i(\d_\m^{\ \r}\d_\n^{\ \s}-\d_\m^{\ \s}\d_\n^{\ \r}),
\ee
where $k_\mn$ are momenta conjugated to the $x_\mn$,
and define $X_\mn=(S\,x\,S^T)_{\m\n}$ and $K^\mn=(S^\ddagger\,k\,S^\mo)^\mn$, where $S^\ddagger=(S^\mo)^T$.

These variables satisfy \cor analogous to (\ref{heis}),
\be\lb{HEIS}
[X_\mn,X_{\r\s}]=[K^\mn,K^{\r\s}]=0,\qquad[X_\mn,K^{\r\s}]=i(\d_\m^{\ \r}\d_\n^{\ \s}-\d_\m^{\ \s}\d_\n^{\ \r}),
\ee
but their indices are raised and lowered by the metric $g_\mn$, for example $K_\mn=g_{\m\r}g_{\n\s}K^{\r\s}$.

The variables $X_\mn$ can be decomposed as
\be\lb{dr1}
X_{i4}={X_i\over\b},\quad X_{i5}={P_i\over\a},\quad X_{45}={H\over\a\b}.
\ee
Moreover, the explicit relations between the four-dimensional components of $K_\mn$ and $k_\mn$ can be written as
\bea\lb{K-k}
&&K_\ij=k_\ij+{\b\over\s}(a_iq_j-a_jq_i)+{\a\over\t}(b_iy_j-b_jy_i)-{\a\u\over\s\t}(a_iy_j-a_jy_i)+\a\b(a_ib_j-a_jb_i)w,\cr
&&\cr
&&Q_i={1\over\s\t}\left(\u q_i-{\a\t\over\b}y_i+\a b_iw\right),\cr
&&\cr
&&Y_i={1\over\s\t}(\s y_i-\a a_iw),\cr
&&\cr
&&W={w\over\s\t},
\eea
with inverse
\bea
&&k_\ij=K_\ij-\b(a_iQ_j-a_jQ_i)-\a(b_iY_j-b_jY_i)+\ab(a_ib_j-a_jb_i)W,\cr
&&\cr
&&q_i=\s(Q_i-\a b_iW)+\u\left({\a\over\b}Y_i+\a a_iW\right),\cr
&&\cr
&&y_i=\t(Y_i+\b a_iW),\cr
&&\cr
&&w=\s\t W,
\eea
where we have defined $k^{i4}=\b q^i$, $k^{i5}=\a y^i$, $k^{45}=\a\b w$ and
\be\lb{dr2}
K^{i4}=\b Q^i,\quad K^{i5}=\a Y^i,\quad K^{45}=\a\b W,
\ee
so that
\be
[X_i,Q_j]=i\d_\ij,\qquad[P_i,Y_j]=i\d_\ij,\qquad[H,W]=i.
\ee

We now want to find a realisation of the $\tM_\mn$ defined in sect.~1 in terms of the Heisenberg algebra generated by $X_\mn$ and $K^\mn$.
The $\tM_\mn$ satisfy the $so(1,N;g)$ algebra (\ref{cor}), that we write as
\be\lb{alg}
[\tM_\mn,\tM_{\r\s}]=i\,C_{\mn,\,\r\s}^{\qquad\a\b}\,\tM_{\a\b}.
\ee

The structure constants are given by
\be
C_{\mn,\,\r\s}^{\qquad\ab}=\ha\Big[-g_{\n\r}(\d_\m^{\ \a}\d_\s^{\ \b}-\d_\m^{\ \b}\d_\s^{\ \a})+g_{\m\s}(\d_\r^{\ \a}\d_\n^{\ \b}-\d_\r^{\ \b}\d_\n^{\ \a})
-(\m\lra\n)\Big],
\ee
and obey the symmetry properties
$C_{\mn,\r\s}^{\qquad\ab}=-C_{\n\m,\r\s}^{\qquad\ab}=-C_{\mn,\s\r}^{\qquad\ab}=-C_{\mn,\r\s}^{\qquad\b\a}$ $=-C_{\r\s,\mn}^{\qquad\ab}$.
Note that structure constants in (\ref{alg}) are multiplied by $\hbar$, which is set to 1 in our conventions, and in the classical limit $\hbar = 0$
all generators commute.

In general, if the operators $M_\mn$ generate a Lie algebra with structure constants $C_{\mn,\r\s}^{\qquad\ab}$,
the universal realization of $M_\mn$ in terms of the Heisenberg algebra (\ref{HEIS}), corresponding to Weyl-symmetric ordering,
is given by \cite{MMK,MSK}
\be\lb{weyl}
M_\mn=X_\ab\left[{{\cal C}\over 1-e^{-\cal C}}\right]_{\mn}^{\quad\ab},
\ee
where $\cC_{\mn}^{\quad\ab}=-\ha\,C_{\mn,\,\r\s}^{\qquad\ab}K^{\r\s}$.

This realization enjoys the property
\be
e^{{i\over2}t^\mn\tM_\mn}\triangleright1=e^{{i\over2}t^\mn X_\mn},
\ee
where the $t_\mn$ are real numbers transforming as tensors under $so(1,N;g)$ and the action $\triangleright$ is defined as
\be
X_\mn\triangleright f(X_\ab)=X_\mn f(X_\ab),\qquad K^\mn\triangleright f(X_\ab)=-i{\de f(X_\ab)\over\de X_\mn}=[K^\mn, f(X_\ab)].
\ee
In particular,
\bea
&&X_\mn\act=X_\mn,\qquad K_\mn\act=0,\cr&&\cr
&&K^\mn\triangleright e^{{i\over2}t^{\a\b}X_{\a\b}}=t^\mn e^{{i\over2}t^{\a\b}X_{\a\b}}
\eea

We can now expand (\ref{weyl}) in powers of the structure constants $C_{\mn,\r\s}^{\qquad\ab}$ (\ie in terms of $\hbar$),
Then the Weyl realization of $\tM_\mn$ in terms of the generalized Heisenberg algebra generated by $X_\mn$ and $K^\mn$ reads
up to second order,
\be\lb{Mrep}
\tM_\mn=X_\mn+{1\over2}X_\ab\,{\cal C}_{\mn}^{\quad\ab}+{1\over12}X_\ab\left({\cal C}^2\right)_{\mn}^{\quad\ab}.
\ee
where
\bea
&&{\cal C}_{\mn}^{\quad\ab}=\ha\Big(\d_\m^{\ \a}K_\n^{\ \b}+\d_\n^{\ \b}K_\m^{\ \a}-(\a\lra\b)\Big),\cr
&&\left({\cal C}^2\right)_\mn^{\quad\ab}=\ha\Big(2K_\m^{\ \a}K_\n^{\ \b}+\d_\n^{\ \b}K_{\m\r}K^{\r\a}+\d_\m^{\ \a}K_{\n\r}K^{\r\b}-(\a\lra\b)\Big),
\eea
and the indices are lowered by means of the metric $g_\mn$.

Inserting $\cal C$ in (\ref{Mrep}), we find up to first order,
\be\lb{real}
\tM_\mn=\ X_\mn+{1\over 2}\left(X_{\m\a}K_\n^{\ \a}-X_{\n\a}K_\m^{\ \a}\right),
\ee
and
\be
[\tM_\mn,K^{\r\s}]=i(\d_\m^{\ \r}\d_\n^{\ \s}-\d_\m^{\ \s}\d_\n^{\ \r})+{i\over2}(\d_\m^{\ \r}K_\n^{\ \s}-\d_\n^{\ \r}K_\m^{\ \s}+\d_\n^{\ \s}K_\m^{\ \r}-\d_\m^{\ \s}K_\n^{\ \r}).
\ee
We can write (\ref{real}) in terms of four-dimensional variables, defined by (\ref{dr}), (\ref{dr1}) and (\ref{dr2}), as
\bea
&&\tM_\ij=X_\ij+\ha\Big(X_{ik}(K_j^{\ k}-\b a_jQ^k-\a b_jY^k)+X_i(Q_j-\a b_jW)+P_i(Y_j+\b a_jW)\cr
&&\qquad\ -(i\lra j)\Big),\cr
&&\cr
&&\tx_i=X_i+\ha\Big(-\b X_\ij(a_kK^{jk}+\b\cA Q^j+\a\tr Y^j)+\b X_i(a_jQ^j-\a\tr W)\cr
&&\qquad+X_j(K_i^{\ j}-\b a_iQ^j-\a b_iY^j)+\b P_i(a_jY^j+\b \cA\,W)-H(Y_i+\b a_i W)\Big),\cr
&&\cr
&&\tp_i=P_i+\ha\Big(-\a X_\ij(b_kK^{jk}+\a \cB Y^j+\b\tr Q^j)+\a P_i(b_jY^j+\b\tr W)\cr
&&\qquad+P_j(K_i^{\ j}-\a b_iY^j-\b a_iQ^j)+\a X_i(\b b_jQ^j-\a\cB\,W)+H(Q_i-\a b_i W)\Big),\cr
&&\cr
&&\tH=H+\ha\Big(\a X_i(\a\cB Y^i-\b\tr Q^i+b_jK^\ij)-\b P_i(\b\cA Q^i+\a\tr Y^i+a_jK^\ij)\cr
&&\qquad\ +H(\b a_iQ^i+\a b_iY^i)\Big),
\eea
where Latin indices are lowered using the flat metric.

Other realizations can be obtained from the Weyl realization $\tM_\mn^W$ using similarity transformations of the type
\be
\tM_\mn= S\,\tM^W_\mn\,S^\mo
\ee
where $S=\exp(G)$ with $G$ of the form $G=X F(K)$. Then the corresponding realizations will be linear in $X$ and can be written as series in $K$.
The corresponding coproducts, star products and twists can be obtained using the same similarity transformations \cite{MS}.

\section{Coproduct and star product in Weyl realization}
Formulae for coproduct and deformed addition of momenta can be deduced using the results of \cite{kSny,MSK}, with the difference that now the sums
are performed with the curved metric $g_\mn$ instead of the flat metric.
This formalism allows us to construct a coproduct for the (\kd) Yang model, in particular, both the momenta conjugated to $\tx_i$ and to $\tp_i$
will admit a coproduct.

Defining
\be
e^{{i\over2}s^{\mn}\tM_{\mn}} e^{{i\over2}t^{\r\s}\tM_{\r\s}}=e^{{i\over2}(s^{\mn}\oplus t^\mn)\tM_{\mn}}\id e^{{i\over2}\cD^\mn(s,t)\tM_\mn},
\ee
where $s^\mn$ and $t^\mn$ transform as $so(1,5;g)$ tensors, one has at first order
\be\lb{Di}
\cD^\mn(s^\ab,t^\ab)=\ s^\mn+t^\mn-{1\over2}\Big(s^{\m\a}t^\n_{\ \a}-s^{\n\a}t^\m_{\ \a}\Big).
\ee
In the following, we shall write all the formulas up to first order, without explicitly mentioning it.

The coproduct $\D K^\mn$ is then
\be\lb{cop}
\D K^\mn=\cD^\mn(K^\mn\otimes1,1\otimes K^\mn)=\ \D_0K^\mn-\ha\Big(K^{\m\a}\otimes K^\n_{\ \a}-K^{\n\a}\otimes K^\m_{\ \a}\Big),
\ee
where $\D_0K^\mn=K^\mn\otimes1+1\otimes K^\mn$.
The coproduct (\ref{cop}) is coassociative.

In components, it reads
\bea
\D K^\ij&=&\D_0 K^\ij-{1\over2}\Big(K^{ik}\ot K^j_{\ k}+\b^2\cA Q^i\ot Q^j+\a^2\cB Y^i\ot Y^j\cr
&&+\a\b\tr(Q^i\ot Y^j+Y^i\ot Q^j)+\b a_k(K^{ik}\ot Q^j+Q^i\ot K^{jk})\cr
&&+\a b_k(K^{ik}\ot Y^j+Y^i\ot K^{jk})-(i\lra j)\Big),\cr
&&\cr
\D Q^i&=&\D_0 Q^i-{1\over2}\Big(K^{ik}\ot Q_k-Q_k\ot K^{ik}+\a\b\tr(Q^i\ot W-W\ot Q^i)\cr
&&+\a^2\cB(Y^i\ot W-W\ot Y^i)+\a b_k(K^{ik}\ot W-W\ot K^{ik})\Big),\cr
&&\cr
\D Y^i&=&\D_0 Y^i-{1\over2}\Big(K^{ik}\ot Y_k-Y_k\ot K^{ik}-\a\b\tr(Y^i\ot W-W\ot Y^i)\cr
&&+\b^2\cA(Q^i\ot W-W\ot Q^i)-\b a_k(K^{ik}\ot W-W\ot K^{ik})\Big),\cr
&&\cr
\D W&=&\D_0W-{1\over2}\Big(Q^k\ot Y_k-Y^k\ot Q_k\Big).
\eea
Using the relations (\ref{K-k}), the coproduct can also be written in terms of tensors transforming under $so(1,5)$.
It is also easy to see that the antipodes are trivial.
\medskip

The star product can easily be deduced from the previous relations, since it is defined as
\be
e^{{i\over2}s^\mn X_\mn}\star e^{{i\over2}t^{\r\s}X_{\r\s}}=e^{{i\over2}\cD^\mn(s,t)X_\mn},
\ee
with $\cD^\mn(s,t)$ given in (\ref{Di}). This star product is associative.

It may be useful to explicitly write down the four-dimensional expression of $\cD^\mn(s,t)$:
setting $\cD^i=\cD^{i4}$, $\bar\cD^i=\cD^{i5}$, $\cD=\cD^{45}$, one has
\bea
\cD^\ij(s,t)&=&\ s^\ij+t^\ij-{1\over2}\Big(s^{ik}t^j_{\ k}+\b^2\cA s^it^j+\a^2\cB\bs^i\bt^j+\a\b\tr(s^i\bt^j+\bs^it^j)\cr
&&+\ \b a_k(s^{ik}t^j+s^it^{jk})+\a b_k(s^{ik}\bt^j+\bs^it^{jk})-(i\lra j)\Big),\cr
&&\cr
\cD^i(s,t)&=&\ s^i+t^i-{1\over2}\Big(s^{ik}t_k-t^{ik}s_k+\a\b\tr(s^it-st^i)+\a^2\cB(\bs^it-s\bt^i)\cr
&&+\ \a b_k(s^{ik}t-st^{ik})\Big)\cr
&&\cr
\bar\cD^i(s,t)&=&\ \bs^i+\bt^i-{1\over2}\Big(s^{ik}\bt_k-\bs_kt^{ik}-\a\b\tr(\bs^it-s\bt^i)+\b^2\cA(s^it-st^i)\cr
&&-\ \b a_k(s^{ik}t-st^{ik})\Big)\cr
&&\cr
\cD(s,t)&=&\ s+t-{1\over2}\Big(s^k\bt_k-\bs^kt_k\Big),
\eea
where we have defined the components of the $so(1,5;g)$ tensors $t^\mn$ as $t^i=t^{i4}$,
$\bar t^i=t^{i5}$, $t=t^{45}$ and analogously for $s^\ij$.

\section{The twist for the Weyl realization}
In this section, we construct the twist operator at first order, again using the results of \cite{kSny,MSK}.
The twist is defined as a bilinear operator such that $\D m=\cF\D_0m\cF^\mo$ for each $m$ belonging
to $so(1,5;g)$.

The twist in a Hopf algebroid sense can be computed by means of the formula \cite{twist}
\be
\cF^\mo\id e^F=e^{-{i\over2}K^\mn\ot X_\mn}e^{{i\over2}K^{\r\s}\ot\tM_{\r\s}}.
\ee
Using the CBH formula one gets
\be\lb{twist}
F={i\over2}\,K^{\m\n}\ot(\tM_\mn-X_\mn).
\ee
and substituting (\ref{real}) in (\ref{twist}), one obtains
\be
F={i\over2}K^{\a\g}\ot X_{\a\b}K_\g^{\ \b}.
\ee

It is easy to check that
\be
\cF\D_0K^\mn\cF^\mo=\D K^\mn,
\ee
with $\D K^\mn$ given in (\ref{cop}).

In terms of components, one can write
\bea
F&=&K^\ij\ot\Big[X_{ik}(K_j^{\ k}-\b a_jQ^k-\a b_jY^k)+X_i(Q_j-\a b_jW)+P_i(Y_j+\b a_jW)\Big]\cr
&&+\ Q^i\ot\Big[-\b X_\ij(\b\cA Q^j+\a\tr Y^j+a_kK^{jk})+\b X_i(a_jQ^j-\a\tr W)\cr
&&+\ X_j(K_i^{\ j}-\b a_iQ^j-\a b_iY^j)+\b P_i(a_jY^j+\b \cA\,W)-H(Y_i+\b a_i W)\Big]\cr
&&+\ Y^i\ot\Big[-\a X_\ij(\a \cB Y^j+\b\tr Q^j+b_kK^{jk})+\a P_i(b_jY^j+\b\tr W)\cr
&&+\ P_j(K_i^{\ j}-\a b_iY^j-\b a_iQ^j)+\a X_i(\b b_jQ^j-\a\cB\,W)+H(Q_i-\a b_i W)\Big]\cr
&&+\ W\ot\Big[\a X_i(b_jK^\ij+\a\cB Y^i-\b\tr Q^i)-\b P_i(\b\cA Q^i+\a\tr Y^i+a_jK^\ij)\cr
&&+\ H(\b a_iQ^i+\a b_iY^i)\Big].
\eea
\section{Coproduct and twist for the original Yang model}

Of course the Hopf structure for the original Yang model can be derived from the previous results simply setting $A=B=1$, $\f=\q=0$ and $a=b=0$.
Since these results are not discussed in the literature we briefly report them here.

The coproduct can be written in terms of the four-dimensional variables defined in section 3 as
\bea
\D K^\ij&=&\D_0 K^\ij-{1\over2}\Big(K^{ik}\ot K^j_{\ k}+\b^2 Q^i\ot Q^j+\a^2 Y^i\ot Y^j\cr
&&+\a\b(Q^i\ot Y^j+Y^i\ot Q^j)-(i\lra j)\Big),\cr
&&\cr
\D Q^i&=&\D_0 Q^i-{1\over2}\Big(K^{ik}\ot Q_k-Q_k\ot K^{ik}+\a\b(Q^i\ot W-W\ot Q^i)\cr
&&+\a^2(Y^i\ot W-W\ot Y^i)\Big),\cr
&&\cr
\D Y^i&=&\D_0 Y^i-{1\over2}\Big(K^{ik}\ot Y_k-Y_k\ot K^{ik}-\a\b(Y^i\ot W-W\ot Y^i)\cr
&&+\b^2(Q^i\ot W-W\ot Q^i)\Big),\cr
&&\cr
\D W&=&\D_0W-{1\over2}\Big(Q^k\ot Y_k-Y^k\ot Q_k\Big).
\eea
It is evident that one cannot disentangle the various components of the conjugated momenta.

Analogously, the twist takes the form
\bea
F&=&K^\ij\ot\Big[X_{ik}K_j^{\ k}+X_iQ_j+P_iY_j\Big]\cr
&&+\ Q_i\ot\Big[-\b X_{ik}(\b Q^k+\a Y^k)+X_kK_i^{\ k}+\b(\b P_i-\a X_i)W-HY^i\Big]\cr
&&+\ Y_i\ot\Big[-\a X_{ik}(\a Y^k+\b Q^k)+P_kK_i^{\ k}+\a(\b P_i-\a X_i)W+HQ^i\Big]\cr
&&+\ W\ot\Big[\a X_k(\a Y^k-\b Q^k)-\b P_k(\b Q^k+\a Y^k)\Big].
\eea
\section{Conclusions}
The Yang model represents a noncommutative geometry defined on a curved background, which is interesting because of possible applications
to quantum cosmology and for its dual nature for the interchange of positions and momenta.

In this paper we have discussed a generalization of that model, called doubly $\k$-deformed Yang model, originally proposed in \cite{LMMPW}
performing a deformation of the flat metric appearing in the definition of the Yang algebra, so that both the de Sitter
 and Snyder subalgebras are
deformed.

The formalism introduced in this paper, inspired to the one used in ref.~\cite{BM} for the Snyder model, permits to define an associative
star product and a coassociative coproduct, together with a twist.
To our knowledge, a Hopf algebra structure for the Yang model has never been discussed before in the literature.
Using this formalism, we have been able to calculate in a straightforward way several properties of the associated Hopf algebra.
The achievement of these results necessitates the introduction  as primary fields of extended tensorial coordinates and a scalar coordinate,
besides position and momentum. In addition also the momenta conjugated to these variables must be considered.

The problem of the interpretation of all these extended coordinates is crucial.
One possibility is that they derive from the symmetry breaking of an $so(1,5,g)$ algebra to an $so(1,3)$ algebra as proposed in
\cite{LMMPW}. Also the possibility to relate Yang models to Kaluza-Klein theories, interpreting the extra degrees of freedom as higher
dimensions is under investigation.

At this point, the question of what are the physical consequences of such models and the possible physical predictions of new observable effects
that could be measured arises. A first step in this direction would be to define a dynamics for the theory, writing down a suitable Hamiltonian
for particles leaving in Yang spacetime. Further developments may include the definition of a quantum field theory compatible with this structure.

\section*{Acknowledgements}
We thank Jerzy Lukierski for useful discussions.
S. Mignemi acknowledges support of Gruppo Nazionale di Fisica Matematica.

\end{document}